\documentclass[12pt]{article}

\usepackage{scicite}
\usepackage{graphicx}
\usepackage{times}

\newenvironment{sciabstract}{%
\begin{quote} \bf}
{\end{quote}}

\title{Changes in the flagellar bundling time account for variations in swimming behavior of flagellated bacteria in viscous media}

\author
{Zijie Qu,$^{1\ast}$ Fatma Zeynep Temel,$^{1,2}$ Rene Henderikx$^{1,3}$, Kenneth S. Breuer$^{1}$\\
\normalsize{$^{1}$School of Engineering, Brown University, Providence, RI 02912, USA}\\
\normalsize{$^{2}$School of Engineering and Applied Sciences, Harvard University, Cambridge, MA 02138, USA}\\
\normalsize{$^{3}$Microsystems group, Eindhoven University of Technology, Netherlands}\\
\\
\normalsize{$^\ast$To whom correspondence should be addressed; E-mail:  zijie\_qu@brown.edu.}
}

\date{}

\begin{document} 




\maketitle


\begin{sciabstract}
Although the motility of the flagellated bacteria, \emph{Escherichia coli}, has been widely studied, the effect of viscosity on swimming speed remains controversial. The swimming mode of wild-type \emph{E.coli} is often idealized as a ``run-and-tumble'' sequence in which periods of swimming at a constant speed are randomly interrupted by a sudden change of direction at a very low speed. Using a tracking microscope, we follow cells for extended periods of time in Newtonian liquids of varying viscosity, and find that the swimming behavior of a single cell can exhibit a variety of behaviors including run-and-tumble and ``slow-random-walk'' in which the cells move at relatively low speed. Although the characteristic swimming speed varies between individuals and in different polymer solutions, we find that the skewness of the speed distribution is solely a function of viscosity and can be used, in concert with the measured average swimming speed, to determine the effective running speed of each cell. We hypothesize that differences in the swimming behavior observed in solutions of different viscosity are due to changes in the flagellar bundling time, which increases as the viscosity rises, due to the lower rotation rate of the flagellar motor. A numerical simulation and the use of Resistive Force theory provide support for this hypothesis. 
\end{sciabstract}


The survival of motile bacteria depends in part on the ability to navigate their environment, swimming towards attractants (e.g. food) and away from repellents (e.g. toxins). In order to move in a low Reynolds number environment and to avoid the time-reversibility of Stokesian dynamics  \cite{purcell1977life}, flagellated bacteria such as \emph{Escherichia coli} exhibit a non-reciprocal swimming behavior first described by Berg and Brown \cite{berg1972chemotaxis}. The ``run-and-tumble'' behavior is characterized by extended linear movements (``runs'') punctuated by sudden changes in direction (``tumbles'').  The tumbling event is initiated by the clockwise (CW) rotation of one or more of the flagellar motors \cite{berg2008coli,turner2000real} ($T_o$ in Fig. 1A). This precipitates the unravelling of the flagellar bundle which causes the cell to immediately stall and re-orient ($T_o - T_1$). As the motor returns to counter-clockwise (CCW) rotation ($T_2$), the flagellar bundle re-forms ($T_2 - T_3$) \cite{turner2000real,kim2003macroscopic} and the cell accelerates back to the characteristic run speed, $U_o$. Note that the value of $U_o$ can vary, and depends on the cell metabolism, the number, length and spatial distribution of flagella and the conditions of the surrounding fluid (temperature, presence or absence of specific nutrients, etc).

This mode of cell motility has been studied extensively over the past decades (e.g. \cite{lighthill1976flagellar,hotani1982micro,berg1993torque,berg1995cells,fraser1999swarming,powers2002role,dalton2011vivo,martinez2014flagellated,patteson2015running}) but while it remains a compelling idealized model for multi-flagellated motion, there remain questions.  For example, 
Molaei \emph{et al.} analyzed thousands of individual cell motion histories  \cite{molaei2014failed} and reported that only $70$\% of the \emph{E.coli} cells exhibited run-and-tumble style of motion while the rest of the cells, moved in a different mode, termed ``slow-random-walk'' and characterized by a slower average speed and absent clearly defined tumbling events. More recently, a close examination of cell motility and flagellar motion \cite{turner2016visualizing} revealed intermediate states, such as partial unbundling, which also contributed to a wider variety of swimming modalities than the binary ``run'' and ``tumble'' states.

Bacteria live in varied fluid environments that can exhibit viscous and/or viscoelastic properties  \cite{kimsey1990motility}, measurements and calculations of cell motility in these complex fluids have yielded numerous seemingly contradictory results \cite{martinez2014flagellated,patteson2015running,berg1979movement,lauga2007propulsion,fu2007theory,leshansky2009enhanced,liu2011force,spagnolie2013locomotion,dasgupta2013speed}.  Even for cells swimming in (assumed to be) Newtonian polymer solutions of varying viscosity, the picture is unclear. One of the earliest experimental studies in polymeric solutions shows that the tumbling of \emph{E.coli} cells is suppressed and swimming speed is increased even when the polymer concentration is low \cite{berg1979movement}. The authors explain this phenomenon by appealing to the properties of the loose and quasi-rigid polymer network and its interactions with the nanoscale flagellar propulsors.  Magariyama and Kudo proposed a simple model based on Resistive Force Theory (RFT)  \cite{purcell1977life,magariyama2002mathematical}, but modified by the introduction of two apparent viscosities that depend on the length, morphology, and the interaction between polymer molecules \cite{magariyama2002mathematical}. A further complication arises from the observation that the level of biological activity appears to change with the addition of the thickening polymer \cite{martinez2014flagellated}, probably due to the metabolism of small polymer fragments by the bacteria.


In order to fully understand the different swimming modes, cells must be observed for relatively long time periods and in different fluid environments. Two methodologies are commonly described. In most studies, cells are tracked under a stationary microscope platform (e.g.  \cite{martinez2014flagellated,molaei2014failed}) which, though effective and straightforward, only permits tracking for short times as the cells quickly pass through the microscope's field of view and focal plane.  Alternatively, one can track individual cells in three dimensions by physically moving the objective and the microscope stage in real time  \cite{berg1972chemotaxis,liu2014helical,turner2016visualizing}. Although the tracking microscope is inefficient, in terms of the number of observed individuals, the extended tracking time 
permits detailed observation of similarities and differences in the swimming behavior for both a single cell and between individual cells in an identical genetic population.

In this manuscript, in an attempt to understand the different swimming modalities and the role of viscosity on cell motility, we report on the use of tracking microscopy to measure the detailed behavior of wild-type \emph{E.coli} swimming in Newtonian fluids of varying viscosity. 
%
%
%
Different solutions of polymers using two molecular weights were prepared, and cell trajectories in both native and dialyzed polymer solutions were recorded.  

A typical time history of speed and angular change (Fig. 1B, C) shows good qualitative and quantitative agreement with the classic results of Berg and Brown \cite{berg1972chemotaxis}. Using their definition of the run and tumble phases (Fig. 1A), we find that the run time and tumbling frequency are not affected by the fluid properties (Table S1). 
However, close inspection of the time-traces indicates that a single cell exhibits both classical run-and-tumble events as well as periods of extended low-speed swimming or ``slow-random-walk''  \cite{molaei2014failed}.  This is quantitatively reflected by the probability density function (PDF) of swimming speed (Fig. 1D) which shows two peaks; one at high speed, which we associate with the observed run behavior, and a second peak at a lower speed corresponding to the ``slow-random-walk'' behavior.

From these results, we assert that the ``slow-random-walk'' mode of motility is not the result of different cells illustrating different swimming modalities. Rather, over an extended period of time, a single cell can exhibit multiple modes of motility. Indeed, more complex combinations of speed and orientation changes are observed, (e.g. Fig. 1B, $t \approx  0$ s $-$ $4$ s) which might be due to partial unbundling  \cite{turner2016visualizing}.

The shape of the speed ($U$) distribution, in particular the skewness, $K = \overline{(U-\overline{U}})^{3}/\sigma^{3}$, where bar denotes the mean value and $\sigma$ is the sample standard deviation, proves to be a valuable means to quantify differences between swimming behaviors. For the example shown (Fig. 1) a bi-modal PDF ($K = 0.22$) illustrates a co-existence between run-and-tumble and ``slow-random-walk'' behaviors.  One can imagine that a swimmer exhibiting a pure run-and-tumble behavior would have a PDF characterized by a sharp peak at the run speed with a broad low-speed tail.  Such a speed distribution would have a negative skewness ($K < 0$).  Similarly, a cell that spends more time in a tumbling state, with only short runs would have a low mean speed and a positively-skewed PDF ($K > 0$). A cell that tumbles continuously, would have a zero mean speed and zero skewness.  Note that the skewness is independent of the absolute swimming speed.



Every cell trajectory that we have measured exhibits this characteristic twin-peaked speed distribution. However, even though the average run times and tumble frequencies are relatively constant (Table S1), there is considerable cell-to-cell variation in absolute swimming speed (Fig. 2A),  most likely due to natural variations in the length and number of flagella. In addition, we observe that there is a marked difference between the swimming speed in dialyzed and native polymer solutions despite the fact that these solutions have the same bulk viscosity (Fig. S1). Although the average swimming speed does decrease as viscosity rises, there does not appear to be a uniform scaling; in addition the swimming speed in native solutions increases initially, before decaying, a phenomenon that has been previously observed \cite{berg1979movement,magariyama2002mathematical,martinez2014flagellated} and variously attributed to non-Newtonian interactions with the polymer network \cite{magariyama2002mathematical} and the effect of small polymer fragments on the metabolic activity of the cell \cite{martinez2014flagellated}.

Characterizing motility purely by the average swimming speed thus appears to be too blunt a tool; however, looking at the skewness of the speed distributions we see that, as the viscosity increases, the speed distribution changes smoothly reflecting a shift from a predominantly run-and-tumble style, characterized by a negative skewness, to a predominantly slow-random-walk style of swimming, characterized by a skewness close to zero, or even slightly positive (Fig. 2B). The same behavior is observed in all four polymer solutions (two different molecular weights, dialyzed and native solutions) suggesting that the speed distribution is consistent with the solution viscosity and independent of the absolute swimming speed.

What might be the cause of this change in the swimming speed distribution?  Assuming that the geometry of the cell body and flagellar filaments do not depend on the fluid viscosity, the hydrodynamics of the run scale linearly with viscosity \cite{purcell1977life, lauga2009hydrodynamics}. Furthermore, the tumbling frequency, 
is independent of viscosity (Table S1).  However, the time for the flagellar bundle to unravel and reform during the tumble \emph{does} change with viscosity. Kim \emph{et al.}  \cite{kim2003macroscopic} showed that the flagellar bundling of elastic helices depends on a non-dimensional parameter, $M = \mu \omega L^4/ E I$, where $\mu$ is the fluid viscosity, $\omega$ is the rotation rate, $L$ the filament length, $E$ the elastic modulus and $I$ the moment of inertia. $M$ represents the balance between the viscous and elastic stresses in the filament and Kim \emph{et al.} demonstrated that flagellar bundling occurs after about $15$ rotations for values of $M$ greater than about $100$. 
For a fixed torque motor \cite{chen2000torque,darnton2007torque}, the flagella rotation rate will decrease as the fluid viscosity increases indicating that the bundling process, which requires a fixed number of rotations \cite{kim2003macroscopic} will take longer at higher viscosity. In addition, Turner \emph{et al.} observed that the swimming speed of the cell remains depressed after tumble due to the rebundling process \cite{turner2000real}. Thus it seems plausible that, as the viscosity rises, the cell spends less time running at full speed, and more time at lower speed recovering from tumbles.  This hypothesis is consistent with our observation that the speed distribution skewness approaches or passes zero as the viscosity rises (Fig. 2B). The effect is obscured in the speed vs. viscosity data (Fig. 2A) by the confounding factors of individual variations in morphology and metabolic activity as well as the effects of polymers on cell activity level.


A numerical simulation confirms the relationship between the bundling time, the scaled average speed, and the skewness of the speed distribution. We model the swimming as a combination of a run at a given ``characteristic run speed'', $U_o$, punctuated by tumbles that occur randomly according to a Poisson distribution. The acceleration from the tumble back to $U_o$ is changed by the effect of varying viscosity on the bundling dynamics. Using this idealized simulation, we generate synthetic trajectories, speed histories and speed distributions associated with different bundling times (Fig. 3A, B) that are both qualitatively and quantitatively similar to the experimentally-measured distributions (e.g. Fig. 1D).  Plotting the distribution skewness against the ratio of the average speed to the characteristic run speed, $\overline{U}/U_o$, we find that the data exhibits a linear trend: $\overline{U}/U_o = -0.185K + 0.627$.
More importantly, the simulation results allow us to use measurements of the speed distribution skewness, $K$, and the average swimming speed, $\overline{U}$, to estimate the characteristic run speed, $U_o$ - a parameter that varies from cell to cell and is not directly measurable. With the estimate for $U_o$, and typical values for the geometry of the cell and flagella \cite{magariyama2002mathematical}, we use RFT to calculate the motor torque, $T$, as well as the cell and flagellar rotation rate, $\omega_c$ and $\omega_f$ respectively.  

Although there is scatter in the data, the  motor torque is estimated to lie between $0.25$ and $0.75$ $\times 10^{-18}$ Nm (Fig. 4A), which agrees well with the measurement of Darnton \emph{et al.} \cite{darnton2007torque} who used a similar technique, but is lower than the measurement of Reid \emph{et al.} \cite{reid2006maximum}. It is worthwhile to note that the motor torque in the native polymer solutions (circles in Fig. 4A) is higher than the torque in the dialyzed solutions (squares in Fig. 4A), consistent with the observations both here, and by Martinez \emph{et al.} \cite{martinez2014flagellated} that the cell activity is generally higher in the native polymer solutions.

Using the motor torque and flagellar rotation rate obtained from RFT, we calculate the bundling time, $T_b$ (Fig. 4B), assuming that $20$ rotations are required for complete bundling.  The results confirm the hypothesis that the bundling time is a function of viscosity, rising from approximately $0.1$ seconds in pure motility buffer to about $0.2$ seconds in the most viscous medium.

A second, independent, estimate of the bundling time can be found from the measured speed vs. time history of each cell. To accomplish this, we first use the skewness of the measured speed distribution to determine the characteristic run speed, $U_o$ (Fig. 3C). Using Berg \& Brown's definition of a change in angular orientation greater than $35^\circ/0.08$ s we identify the start of each tumble ($T_o$ in Fig. 1A) and mark the completion of the re-bundling process as the time at which the swimming speed first reaches the characteristic run speed ($T_3$ in Fig. 1A).  The bundling time, $T_b$, is then defined as  $T_3 - T_o - 0.32$, where $0.32$s is used as the duration of the CW rotation ($T_2 - T_o$) (Fig. 1A) \cite{berg2008coli,turner2000real,darnton2007torque}.

The estimate of the flagellar bundling time obtained using this method (Fig. 4C) agrees well with the results obtained using RFT (Fig. 4B), demonstrating that the bundling time increases with viscosity, rising from about $0.08$ to $0.3$ seconds over the five-fold increase in viscosity. The scatter in the data likely results from our inability to accurately estimate the exact duration of the CW rotation ($T_2 - T_o$) and the variability associated with the determination of $T_3$.  

In summary, we confirm that the motility of a wild-type \emph{E. coli} cell is quite nuanced, exhibiting both run-and-tumble and slow-random-walk during its natural swimming behavior. Furthermore, we believe that we have clarified the confusion surrounding cell motility in viscous media by demonstrating that the swimming behavior in Newtonian fluids of different viscosities can be explained using classical Resistive Force Theory coupled with the recognition that the flagellar bundling process takes longer at higher viscosity due to the slower rotation of the flagellar motor. Lastly, we note that the skewness of the swimming speed is a more useful metric than the average swimming speed, and can be used in the analysis of cell swimming trajectories to control for differences in the characteristic run speed that arise due to cell-to-cell variations and for differences related to the uncontrolled presence of biological stimulants in the surrounding medium.

\newpage

\clearpage
\begin{figure*}
\centering
\includegraphics[width=1\linewidth]{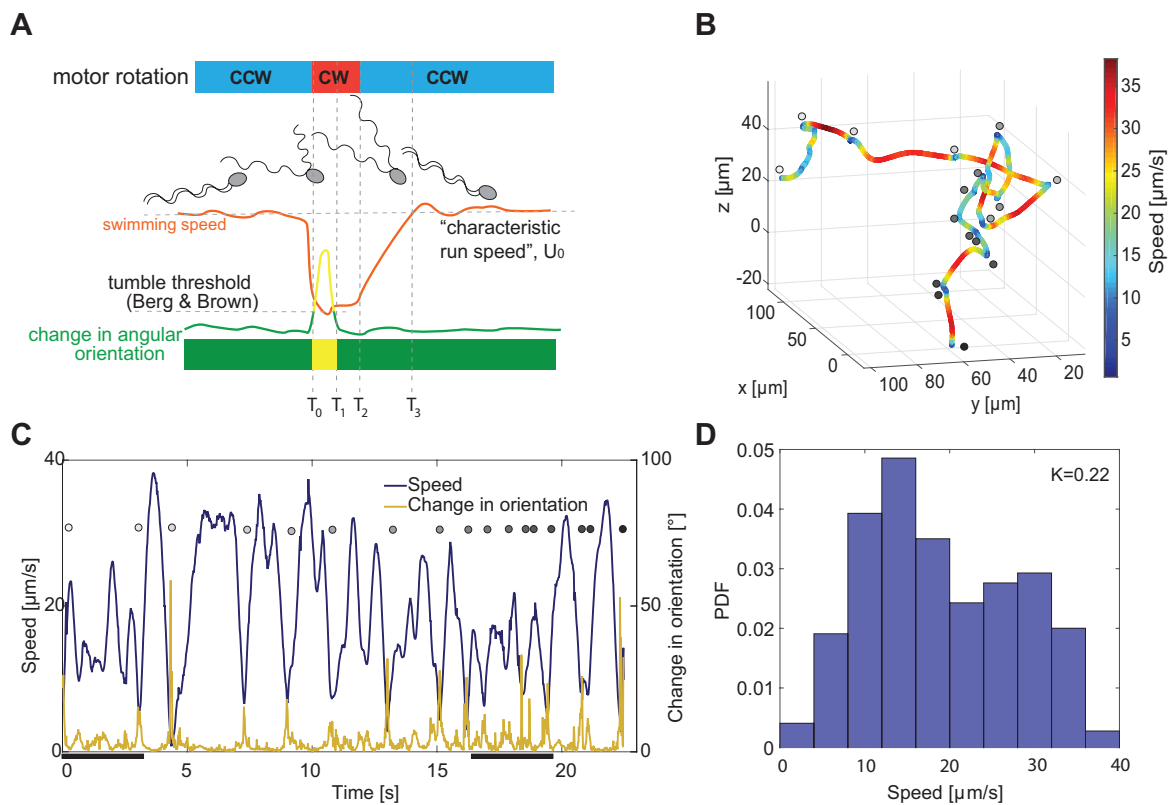}
\caption{Demonstration of diverse swimming modalities. 
(A) Schematic of the tumbling process, (adapted from Darnton \cite{darnton2007torque}). $T_o$: the initiation of tumble when motor starts to rotate CW; $T_{1}$: the end of tumble according to the definition given by Berg and Brown \cite{berg1972chemotaxis}; $T_{2}$: the motor starts to rotate CCW and the re-bundling is initiated; $T_{3}$: the completion of bundle process when the swimming speed reaches the characteristic run speed $U_o$. 
%
(B) Three-dimensional trajectory of a representative \emph{E.coli} cell swimming in 1.25\% Ficoll 400 solution ($1.17$ cP); color change denotes the speed of the cell. 
(C) Time history of swimming speed (blue) and change in orientation (yellow); the round markers on both (B) and (C) denote a tumble event using the definition of Berg and Brown  \cite{berg1972chemotaxis}.  Markers with the same color refer to the same event. The black bars on the $x$-axis of (C) identify  periods of ``slow-random-walk'' \cite{molaei2014failed}. 
(D) The corresponding probability distribution function of the swimming speed;  the two peaks at $12$ $\mu$m/s and $30$ $\mu$m/s correspond to the slow-random-walk and run motilities.}
\label{fig:fig1}
\end{figure*}

\clearpage
\begin{figure*}
\centering
\includegraphics[width=1\linewidth]{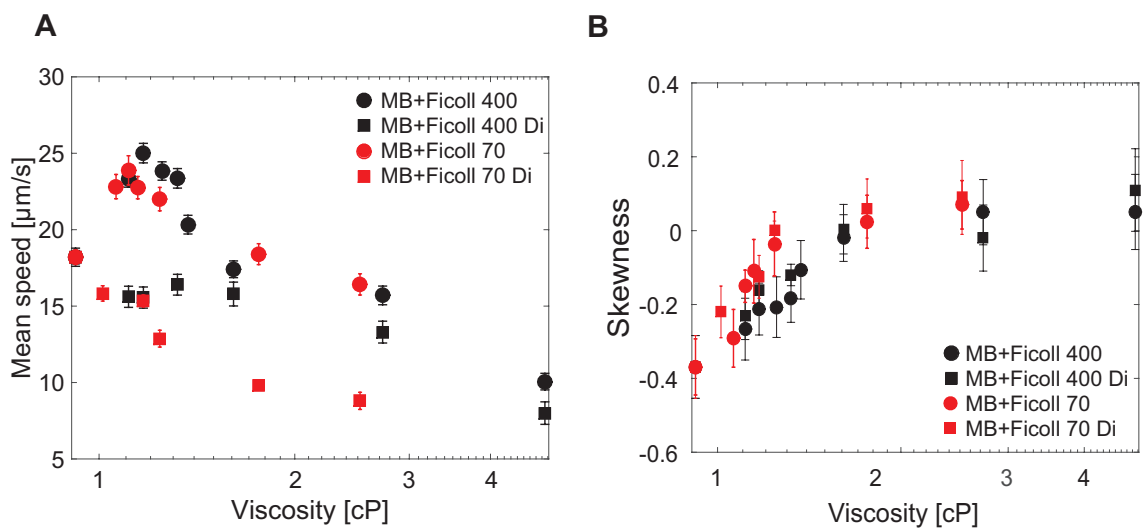}
\caption{(A) Average swimming speed as a function of viscosity for Ficoll 400 and Ficoll 70 solutions (native and dialyzed). (B) Skewness of the swimming speed distribution as a function of viscosity. Although the mean speed exhibits variations as a function of the viscosity and the specific polymer solution, the skewness of the swimming speed distribution demonstrates a  unified behavior, depending only on viscosity.}
\label{fig:2}
\end{figure*}

\clearpage
\begin{figure*}
\centering
\includegraphics[width=1\linewidth]{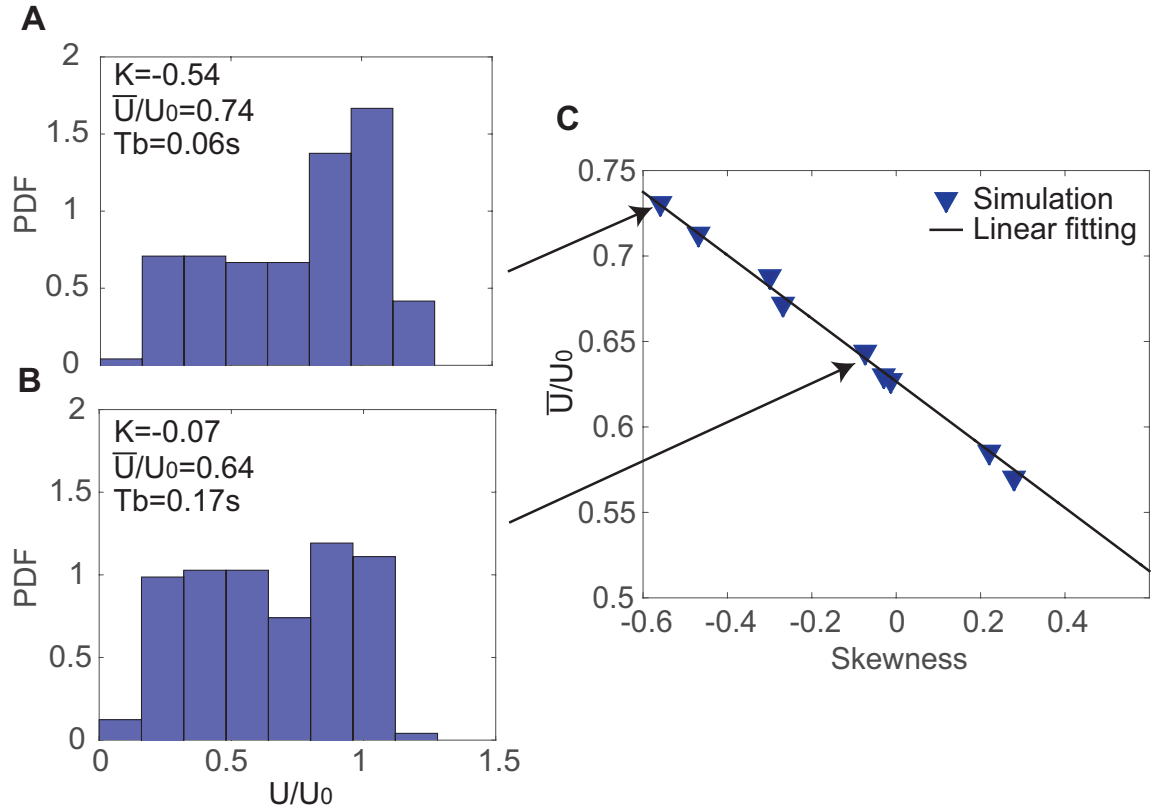}
\caption{Idealized numerical simulations of swimming are defined by a characteristic run speed ($U_o$, held constant at $25$ $\mu$m/s), a tumbling frequency ($1$ Hz) and a bundling time, $T_b$ (varied, to simulate the effects of viscosity on the flagellar bundling process). The distribution of swimming speeds for (A) a ``pure'' run-and-tumble swimmer ($T_b = 0.06$ s, $K = -0.54$), and (B) a combined swimmer ($T_b = 0.17$ s, $K = -0.07$), show the effects of bundling time on the overall distribution.  (C) A linear relationship is observed between the skewness of the swimming speed, $K$, and the ratio of the mean speed to characteristic run speed: $\overline{U}/U_o = -0.185 \times K + 0.627$ (corresponding to $0.06$ s $< T_b < 0.32$ s).}
\label{fig:fig3}
\end{figure*}

\clearpage
\begin{figure*}
\centering
\includegraphics[width=1\linewidth]{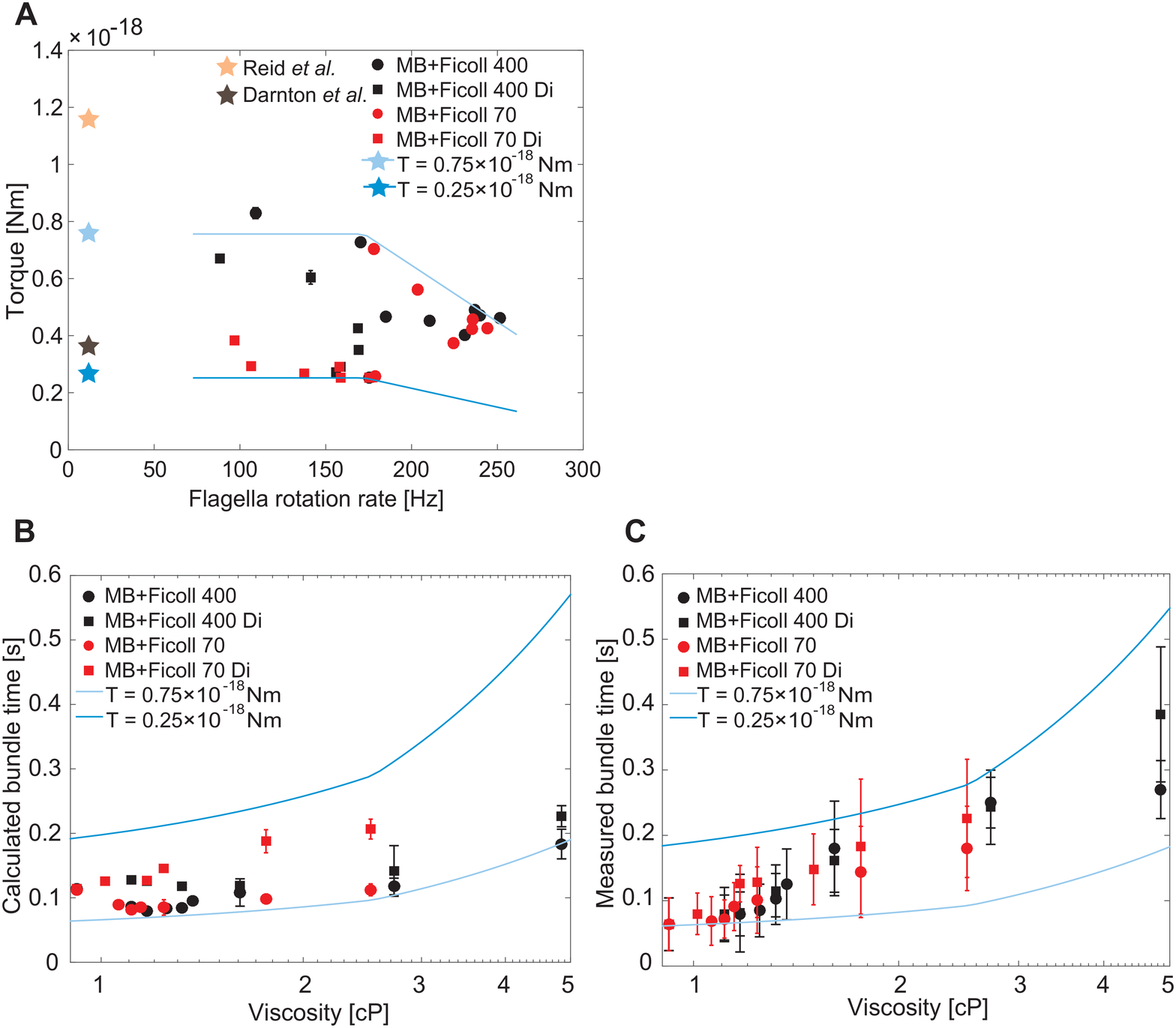}
\caption{
(A) Flagellar motor torque, calculated using RFT and using the measured characteristic run speed, $U_o$, and a typical cell geometry \cite{magariyama2002mathematical}. The blue solid lines are upper and lower bounds  of the torque-speed characteristic, assuming a fixed ``knee" speed at $175$ Hz \cite{chen2000torque}. Previous motor torque measurement by Reid \emph{et al.} \cite{reid2006maximum} and Darnton \emph{et al.} \cite{darnton2007torque} are shown for comparison. 
(B) Calculated bundling time, $T_b$, as a function of viscosity. Here the calculations are based on RFT, using the characteristic run speed, $U_o$, determined from the skewnewss of the speed distribution from Fig. 3C.
(C) Measured bundling time as a function of viscosity. The bundling time is calculated from the swimming histories.
In both (B) and (C), the solid lines are calculated from RFT using motor torque characteristics given in (A).}
\label{fig:fig4}
\end{figure*}

\clearpage



\bibliography{scibib}

\begin{thebibliography}{10}

\bibitem{purcell1977life}
E.~M. Purcell, {\it Am. J. Phys\/} {\bf 45}, 3 (1977).

\bibitem{berg1972chemotaxis}
H.~C. Berg, D.~A. Brown, {\it et~al.\/}, {\it Nature\/} {\bf 239}, 500 (1972).

\bibitem{berg2008coli}
H.~C. Berg, {\it E. coli in Motion\/} (Springer Science \& Business Media,
  2008).

\bibitem{turner2000real}
L.~Turner, W.~S. Ryu, H.~C. Berg, {\it Journal of bacteriology\/} {\bf 182},
  2793 (2000).

\bibitem{kim2003macroscopic}
M.~Kim, J.~C. Bird, A.~J. Van~Parys, K.~S. Breuer, T.~R. Powers, {\it
  Proceedings of the National Academy of Sciences\/} {\bf 100}, 15481 (2003).

\bibitem{lighthill1976flagellar}
J.~Lighthill, {\it SIAM review\/} {\bf 18}, 161 (1976).

\bibitem{hotani1982micro}
H.~Hotani, {\it Journal of molecular biology\/} {\bf 156}, 791 (1982).

\bibitem{berg1993torque}
H.~C. Berg, L.~Turner, {\it Biophysical journal\/} {\bf 65}, 2201 (1993).

\bibitem{berg1995cells}
H.~C. Berg, L.~Turner, {\it Proceedings of the National Academy of Sciences\/}
  {\bf 92}, 477 (1995).

\bibitem{fraser1999swarming}
G.~M. Fraser, C.~Hughes, {\it Current opinion in microbiology\/} {\bf 2}, 630
  (1999).

\bibitem{powers2002role}
T.~R. Powers, {\it Physical Review E\/} {\bf 65}, 040903 (2002).

\bibitem{dalton2011vivo}
T.~Dalton, {\it et~al.\/}, {\it PloS one\/} {\bf 6}, e27317 (2011).

\bibitem{martinez2014flagellated}
V.~A. Martinez, {\it et~al.\/}, {\it Proceedings of the National Academy of
  Sciences\/} {\bf 111}, 17771 (2014).

\bibitem{patteson2015running}
A.~Patteson, A.~Gopinath, M.~Goulian, P.~Arratia, {\it Scientific reports\/}
  {\bf 5} (2015).

\bibitem{molaei2014failed}
M.~Molaei, M.~Barry, R.~Stocker, J.~Sheng, {\it Physical review letters\/} {\bf
  113}, 068103 (2014).

\bibitem{turner2016visualizing}
L.~Turner, L.~Ping, M.~Neubauer, H.~C. Berg, {\it Biophysical Journal\/} {\bf
  111}, 630 (2016).

\bibitem{kimsey1990motility}
R.~B. Kimsey, A.~Spielman, {\it Journal of Infectious Diseases\/} {\bf 162},
  1205 (1990).

\bibitem{berg1979movement}
H.~C. Berg, L.~Turner, {\it Nature\/} {\bf 278}, 349 (1979).

\bibitem{lauga2007propulsion}
E.~Lauga, {\it Physics of Fluids (1994-present)\/} {\bf 19}, 083104 (2007).

\bibitem{fu2007theory}
H.~C. Fu, T.~R. Powers, C.~W. Wolgemuth, {\it Physical review letters\/} {\bf
  99}, 258101 (2007).

\bibitem{leshansky2009enhanced}
A.~Leshansky, {\it Physical Review E\/} {\bf 80}, 051911 (2009).

\bibitem{liu2011force}
B.~Liu, T.~R. Powers, K.~S. Breuer, {\it Proceedings of the National Academy of
  Sciences\/} {\bf 108}, 19516 (2011).

\bibitem{spagnolie2013locomotion}
S.~E. Spagnolie, B.~Liu, T.~R. Powers, {\it Physical review letters\/} {\bf
  111}, 068101 (2013).

\bibitem{dasgupta2013speed}
M.~Dasgupta, {\it et~al.\/}, {\it Physical Review E\/} {\bf 87}, 013015 (2013).

\bibitem{magariyama2002mathematical}
Y.~Magariyama, S.~Kudo, {\it Biophysical journal\/} {\bf 83}, 733 (2002).

\bibitem{liu2014helical}
B.~Liu, {\it et~al.\/}, {\it Proceedings of the National Academy of Sciences\/}
  {\bf 111}, 11252 (2014).

\bibitem{lauga2009hydrodynamics}
E.~Lauga, T.~R. Powers, {\it Reports on Progress in Physics\/} {\bf 72}, 096601
  (2009).

\bibitem{chen2000torque}
X.~Chen, H.~C. Berg, {\it Biophysical Journal\/} {\bf 78}, 1036 (2000).

\bibitem{darnton2007torque}
N.~C. Darnton, L.~Turner, S.~Rojevsky, H.~C. Berg, {\it Journal of
  bacteriology\/} {\bf 189}, 1756 (2007).

\bibitem{reid2006maximum}
S.~W. Reid, {\it et~al.\/}, {\it Proceedings of the National Academy of
  Sciences\/} {\bf 103}, 8066 (2006).

\end{thebibliography}

\bibliographystyle{Science}

\section*{Acknowledgments}
We are grateful to Coli Genetic Stock Center (Yale University) for strains and advice. We thank A.Tripathi for the help in measuring the sample viscosity. The work was supported by National Science Foundation (CBET\#1336638).
\section*{Supplementary materials}
Materials and Methods\\
Fig. S1\\
Table S1\\

\end{document}